\documentclass{article}

\usepackage{arxiv}

\usepackage[utf8]{inputenc} 
\usepackage[T1]{fontenc}    
\usepackage{hyperref}       
\usepackage{url}            
\usepackage{booktabs}       
\usepackage{amsfonts}       
\usepackage{nicefrac}       
\usepackage{lipsum}		   
\usepackage{graphicx}
\usepackage{natbib}
\usepackage{doi}
\usepackage{amsmath}
\usepackage{color}

\title{An interaction potential method for passive and active dynamics of hyperelastic materials}

\author{ 
    \href{https://orcid.org/0000-0001-8616-269X}{\includegraphics[scale=0.06]{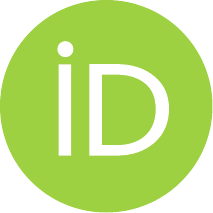}\hspace{1mm}Francesco De Vita}\thanks{francesco.devita@gssi.it} \\
	Gran Sasso Science Institute\\
	Viale F. Crispi 7, L'Aquila, 67100, Italy \\
	\And
	\href{https://orcid.org/0000-0001-5755-4678}{\includegraphics[scale=0.06]{orcid.pdf}\hspace{1mm}Filippo Caruso Lombardi} \\
	Gran Sasso Science Institute\\
	Viale F. Crispi 7, L'Aquila, 67100, Italy \\
	\And
	\href{https://orcid.org/0000-0002-2690-9998}{\includegraphics[scale=0.06]{orcid.pdf}\hspace{1mm}Roberto Verzicco} \\
	Gran Sasso Science Institute\\
	Viale F. Crispi 7, L'Aquila, 67100, Italy \\
	\And
	\href{https://orcid.org/0000-0003-1303-5934}{\includegraphics[scale=0.06]{orcid.pdf}\hspace{1mm}Francesco Viola} \\
	Gran Sasso Science Institute\\
	Viale F. Crispi 7, L'Aquila, 67100, Italy \\
}

\renewcommand{\shorttitle}{Active strain}

\begin{document}
\maketitle

\begin{abstract}
Simulating active biological tissues, such as the myocardium, 
requires constitutive models that are both physically faithful and 
computationally efficient. The most common approach relies on finite element 
methods that accurately discretize the underlying continuum hyperelastic problem 
which, in turn, require global nonlinear solve at each time step. On the other 
hand, fast, interaction-potential methods replace the continuum 
with a network of independent links approximating the mechanical response. 
We propose an interaction potential formulation for 
simulating active biological tissues that bridges this gap. The method recasts 
continuum hyperelastic constitutive laws in terms of tetrahedral edge strain, 
providing an edge-based representation of the element energy. Unlike classical 
mass-spring models, the proposed formulation does not approximate the tissue as 
independent spring elements but preserves the energetic coupling between adjacent 
edges. Passive tissue mechanics is described by hyperelastic constitutive laws, 
while active contraction is incorporated through the active-strain multiplicative 
decomposition. Within the edge-based formulation, the active strain is 
incorporated through a time-dependent activated reference configuration. We 
further introduce a strategy for enforcing the material incompressibility 
constraint while avoiding volumetric locking. The resulting method can be 
interpreted as an edge-strain representation of a constant-strain tetrahedral 
continuum element, providing a bridge between continuum mechanics and discrete 
interaction potential solvers. 
Numerical tests demonstrate the capability of the method to effectively simulate 
different hyperelastic constitutive laws and to properly preserve the energy 
balance equation. Finally, the proposed 
method is applied to the active deformation of a realistic ventricle, reproducing 
longitudinal shortening and wall-thickening values consistent with the literature.

\end{abstract}


\section{Introduction}
The mechanical features of many active biological materials is characterized by 
large deformations and complex responses to electrical activation. Such behavior 
originates from the strong nonlinear coupling between electrophysiological and 
mechanical processes occurring at different spatial scales, as well as from the 
nonlinearity and anisotropy of the materials. Cardiac tissue constitutes a 
representative example. Myocytes, the cardiac muscle cells, are organized in 
bundles of myofibers with their orientation continuously varying throughout the 
ventricular wall, ranging from a right-handed helical arrangement in the 
endocardium to a left-handed helical arrangement in the epicardium 
\cite{nielsen1991mathematical}. Myocytes are elongated cylindrical cells with 
characteristic diameters of 10-20 $\mu$m and lengths of approximately 100 $\mu$m. 
At smaller scales, they contain bundles of myofibrils, which are composed of 
repeating contractile units called sarcomeres, each approximately 2 $\mu$m long.
Cardiac contraction is initiated by electrical excitation. Following 
depolarization of the cell membrane from its resting transmembrane potential 
(approximately $-80$ mV for the left ventricle) to positive values (approximately 
$+20$ mV for the left ventricle), calcium enters 
the cell and triggers actin--myosin cross-bridge formation within the sarcomeres. 
This process generates active force and sarcomere shortening. Subsequently, 
repolarization and calcium removal lead to relaxation of the muscle cell
\cite{bers2002cardiac}. The resulting electromechanical behavior spans multiple 
spatial and temporal scales and involves coupled electrical, biochemical, and 
mechanical processes. Accordingly, computational models must account for action 
potential propagation, excitation--contraction coupling, active contraction, and 
the nonlinear mechanical response of the tissue. In this work, we focus on the 
latter aspect and propose a new method to model the nonlinear mechanical response 
of active biological tissues.

The dynamics of deformable biological tissues is governed by the momentum balance
equation complemented by a constitutive law. The passive response of soft tissues
such as the myocardium is commonly described by hyperelastic constitutive laws, 
in which the stress is derived from a strain-energy density function 
$\mathcal{U}(\mathbf{F})$ of the deformation gradient $\mathbf{F}$, typically 
obtained from the Helmholtz free energy of the system. Several forms have been 
proposed to account for the anisotropy induced by the tissue microstructure, such 
as the transversely isotropic formulation \cite{guccione1995finite} or the 
orthotropic formulation \cite{holzapfel2009constitutive}. Active contraction is 
incorporated through one of two main approaches \cite{ambrosi2012active}: the 
active-stress approach, based on an additive decomposition of the stress into 
passive and active parts, or the active-strain approach, based on a 
multiplicative decomposition of the deformation gradient into an elastic and an 
active component, $\mathbf{F}=\mathbf{F}_e\mathbf{F}_a$ 
\cite{ambrosi2011electromechanical}. Both approaches have been applied to 
simulate cardiac dynamics \cite{goktepe2014generalized,guccione1995finite,
pezzuto2014orthotropic,rossi2014thermodynamically}, and finite element methods 
(FEM) are the reference framework to discretize these models and attain a 
consistent approximation of the underlying continuum problem. Its cost, however, 
stems from solving a large, global system at each step: internal forces are 
obtained from stress measures such as the first Piola--Kirchhoff tensor, and 
implicit schemes require the repeated assembly of stiffness matrices and 
consistent tangent operators together with the solution of large sparse linear 
systems at every Newton iteration and time step \cite{belytschko2014nonlinear,
bonet1997nonlinear}. This global solution step is also the component least suited 
to fine-grained parallelization, being communication-bound, in contrast to the 
purely local evaluation of element contributions. The cost is further increased 
for biological tissues, which are generally nearly incompressible and may suffer
volumetric locking when discretized with low-order elements. This motivates mixed 
displacement--pressure formulations, as well as, stabilized or locking-resistant 
schemes, although at the price of saddle-point systems and inf--sup stability 
requirements \cite{rossi2016implicit,irving2007volume,francu2021locking}.
Anisotropy and, especially, the active-strain decomposition adds a further layer 
of complexity, since $\mathbf{F}_a$ enters the elastic energy through
$\mathbf{F}_e=\mathbf{F}\mathbf{F}_a^{-1}$ and must be propagated consistently 
into the stress and tangent at every quadrature point 
\cite{pezzuto2014orthotropic,garcia2019new}. The outcome is an accurate but 
computationally heavy framework whose cost structure is poorly matched to 
large-scale, real-time, or many-query cardiac simulations.

An attractive alternative for the simulation of deformable soft tissues is the 
interaction potential method \cite{tanaka2012computational,fedosov2010systematic} 
and the particle based  method \cite{macklin2016xpbd}. Within this class of 
methods, the continuum problem is replaced by a discrete mechanical system of 
nodes carrying lumped masses, and the dynamics is advanced by evaluating nodal 
forces from local potentials defined on the mesh connectivity. Because the mass 
matrix is diagonal and the force evaluation is local, these methods avoid the 
solution of large linear systems when advanced explicitly in time, making them 
straightforward to implement and parallelize efficiently, particularly on GPUs. 
The interaction potential method has been used to simulate red blood cells
\cite{fedosov2010systematic,nakamura2013spring}, as well as more complex
systems such as the whole heart \cite{viola2023high,viola2023gpu} and
bat-inspired membrane wings \cite{kumar2025computational}.

In their most common form, however, the material response is assembled from 
independent edge-spring energies, and this is precisely where their physical 
fidelity breaks down: a network of central-force springs does not converge, under 
mesh refinement, to a prescribed continuum constitutive law. The sum of the 
independent spring potentials does not retain the energetic equivalence with the 
continuum strain-energy density. For example, when modeling linear elastic 
materials, an isotropic mass-spring network cannot set the two Lam\'e parameters 
independently and is locked to a fixed Poisson ratio (about 0.33) 
\cite{tanaka2012computational}. Such networks, therefore, cannot represent the 
near-incompressible regime characteristic of most soft tissues, and their 
parameters must be identified by fitting a target response rather than mapped 
from continuum moduli. More elaborate formulations partially close this gap by
augmenting the edge springs with multi-node potentials: area- and 
volume-conservation terms can be included and the discrete parameters can be 
linked to the macroscopic moduli \cite{fedosov2010systematic,nakamura2013spring}.
These remedies, however, must be tailored case by case and do not provide a 
general route from an arbitrary hyperelastic strain energy density 
$\mathcal{U}(\mathbf{F})$ to an equivalent discrete potential.

More recent energy-based formulations have narrowed the distance between fast 
discrete solvers and continuum elasticity. Extended position-based approaches 
such as XPBD \cite{macklin2016xpbd,macklin2021constraint} derive forces from 
elastic energies---including stable Neo-Hookean models---through constraint 
projections. These methods, however, have been developed primarily for real-time 
computer graphics and the constraints are satisfied iteratively with a prescribed 
numerical tolerance.

To the best of our knowledge, no existing interaction-potential formulation 
simultaneously provides \emph{i)}  potentials that are energetically equivalent 
to a prescribed continuum hyperelastic energy at the element level, and 
\emph{ii)} a natural incorporation of the active-strain multiplicative 
decomposition. In the present work we address this gap by rewriting continuum 
hyperelastic element energy directly in terms of the tetrahedral edge strains. 
The discrete energy is therefore not a spring-network surrogate but the continuum
strain-energy density $\mathcal{U}(\mathbf{F})$ itself, evaluated on a discrete 
kinematics defined by the mesh edges. The resulting formulation thus preserves 
the energetic structure of the continuum constitutive law---and with it the 
correct passive response, anisotropy, and Poisson behavior--- while expressing 
the internal forces as local, edge-wise contributions in the form natural to 
interaction-potential solvers, with diagonal mass and explicit time integration. 
Therefore, the method retains the strength of both approaches: the energetic 
consistency of the continuum model and the parallel efficiency of particle-based 
methods. Within this formalism the active-strain decomposition
$\mathbf{F}=\mathbf{F}_e\mathbf{F}_a$ admits a particularly simple realization, 
as a time-dependent activation of the edge reference configuration. We validate 
the method against a set of FEM benchmarks, including the established 
cardiac-mechanics benchmark problems of Land et al. \cite{land2015verification}, 
and show that it attains an accuracy comparable to FEM, represents nearly 
incompressible materials without volumetric locking, and accurately preserves the 
energy balance under active deformation.

The manuscript is organized as follows: section \ref{sec:problem} presents the 
standard interaction potential formulation; section \ref{sec:hyperel} describes 
the new interaction potential formulation for hyperelastic materials both in the 
passive and active dynamics; section \ref{sec:incomp} describes how to model 
nearly incompressible materials avoiding volumetric locking; section 
\ref{sec:numerics} illustrates the numerical implementation of the present 
method; section \ref{sec:validation} presents a collection of numerical 
benchmarks both for the passive ad for the active dynamics; finally section 
\ref{sec:ventricle} shows an application of the proposed method to the active 
deformation of a realistic ventricle.

\section{Interaction potential formulation\label{sec:problem}}

\subsection{General framework\label{sec:problem_general}}
\begin{figure}
\centering
    \includegraphics[width=0.9\textwidth]{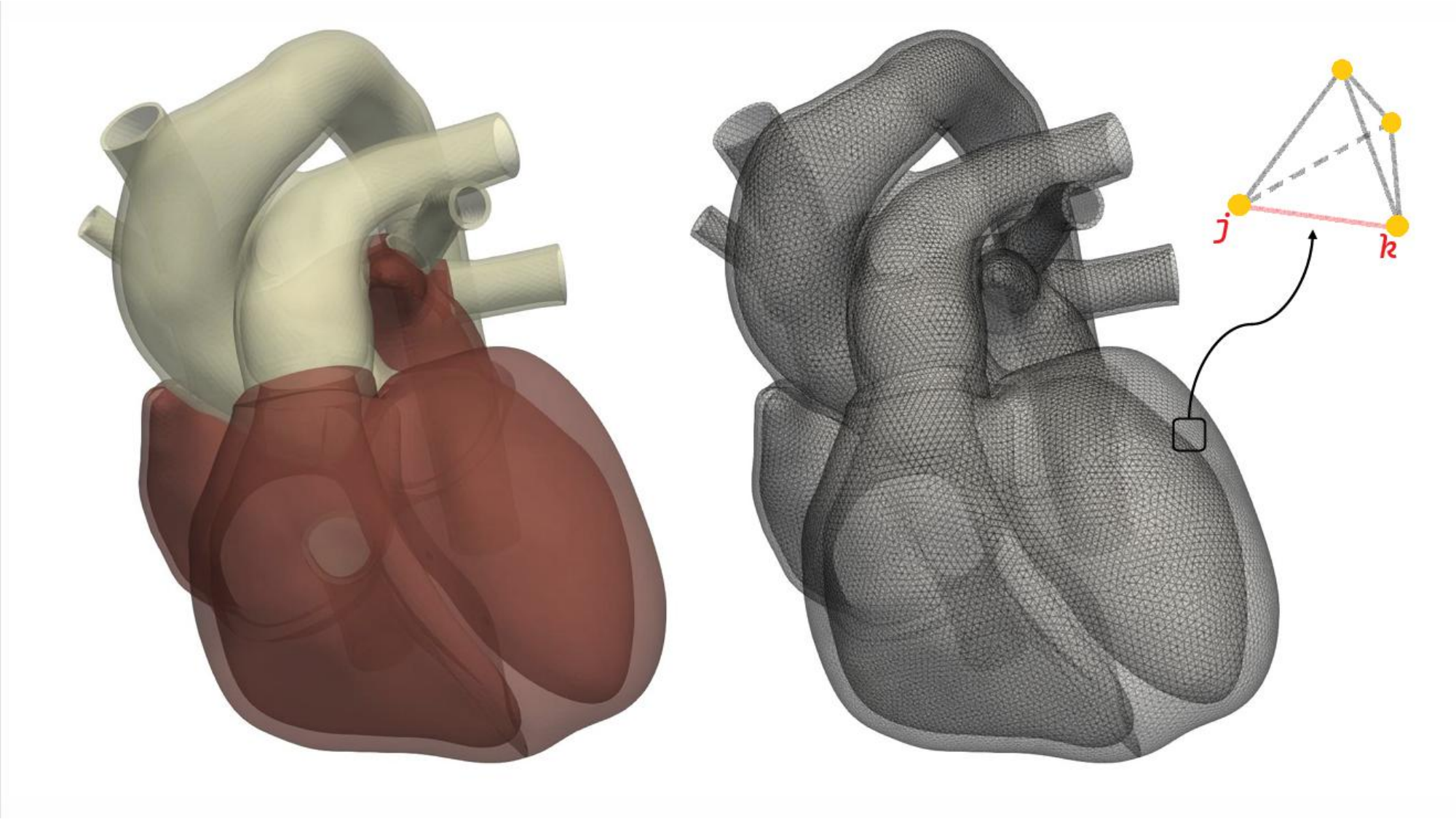}
    \caption{Sketch of the problem: a three-dimensional deformable soft body 
    (such as the human heart) is discretized with tetrahedral elements. The nodes 
    of each tetrahedral cell (yellow dots) represent discrete mass points for 
    which Newton's second law is solved; in red is highlighted one of the six 
    edge of the cell, connecting nodes $j$ and $k$.\label{fig:sketch}}
\end{figure}
The interaction potential method models the deformation of a solid body by 
prescribing potential energies associated with its deformation modes 
\cite{tanaka2012computational,fedosov2010systematic,nakamura2013spring,
viola2023high}. Consider a solid body discretized by $N_c$ tetrahedral cells, 
having $N_v$ vertices (or nodes) connected by $N_e$ edges, as sketched in Figure 
\ref{fig:sketch}. Each vertex represents a discrete mass point, with mass $m_i$ 
and position vector $\mathbf{x}_i$, $i = 1, \ldots, N_v$. The nodal masses are 
obtained by distributing the mass of each cell to its four vertices
\begin{equation}
    m_i = \sum_{c \in \mathcal{E}_c^i} \frac{\rho V_c}{4}
\end{equation}
where $\mathcal{E}_c^i$ is the set of cells incident to node $i$, $V_c$ is the 
cell volume, and $\rho$ is the solid density. Assuming $N_U$ potential energies, 
the resulting internal force acting on node $i$ is given by
\begin{equation}
    \label{eqn:internalf}
    \mathbf{f}_i =  - \sum_{p=1}^{N_U} \frac{\partial U_p}{\partial \mathbf{x}_i}
\end{equation}
and the system dynamics is governed by $N_v$ Newton's second law equations:
\begin{equation}
    \label{eqn:Newton}
    m_i \frac{d^2 \mathbf{x}_i}{d t^2} + \xi \frac{d \mathbf{x}_i}{d t} = 
    \mathbf{f}_i + \mathbf{f}_i^{ext}, \quad i = 1, \ldots N_v
\end{equation}
with $\mathbf{f}_i^{ext}$ the external force acting on the $i$-th mass point and 
$\xi$ a friction coefficient. 

This framework is common to all interaction potential methods, including the one 
proposed in this work (see section~\ref{sec:hyperel}). Two features make this 
approach attractive for large-scale simulations. First, the mass matrix is 
diagonal by construction, so the equations of motion can be advanced explicitly 
without any linear solve. Second, provided that the potentials $U_p$ are local, 
the internal forces \eqref{eqn:internalf} depend only on each node and its mesh 
neighbors, allowing the force evaluation to be efficiently parallelized.
The distinction between interaction potential methods lies not in the discrete 
skeleton itself, but in the choice of the energies $U_p$. This choice determines 
whether the discrete model reproduces a prescribed continuum material behavior or 
only a qualitatively plausible one. The discretization above provides the 
baseline for the proposed formulation. We next review the classical choice of 
$U_p$, the linear-spring network, and its limitations, which motivate our 
approach.

\subsubsection{The linear-spring network\label{sec:problem_springs}}
The simplest choice of interaction potential treats each edge as an independent 
linear spring, with elastic energy:
\begin{equation}
    \label{eqn:linearspring}
    U = \frac{1}{2}k_s \left(l(\mathbf{x}_j, \mathbf{x}_k) - l_0(\mathbf{x}_j, 
    \mathbf{x}_k)\right)^2
\end{equation}
where $l(\mathbf{x}_j, \mathbf{x}_k) = |\mathbf{x}_j - \mathbf{x}_k| = 
|\mathbf{x}_{jk}|$ is the current length of the edge connecting the nodes $j$ and 
$k$, $l_0(\mathbf{x}_j, \mathbf{x}_k) = l(\mathbf{x}_{j}^0,\mathbf{x}_k^0)$ is 
its reference length, and $\mathbf{x}_i^0 = \mathbf{x}_i(t=0)$ is the position 
vector in the stress-free configuration. This potential represents the in-plane 
deformation mode of the material and leads to a classical mass-spring model. The 
connection between the discrete model and the continuum material can be made 
linking each discrete parameter (the spring stiffness) to a macroscopic material 
parameter. For example, a linear elastic material is characterized by its Young's 
modulus and Poisson ratio (or equivalently the Lamé constants). The value of the 
spring stiffness $k_s$ in \eqref{eqn:linearspring} can be tuned to match the 
Young's modulus of the material \cite{gelder1998approximate}; however, this 
results in fixing the Poisson ratio (about 0.33 for an isotropic triangluar mesh 
\cite{tanaka2012computational}). Additional interaction potentials can be 
included to model other macroscopic properties, such as the bending stiffness or 
the bulk modulus of the tissue \cite{tanaka2012computational,
fedosov2010systematic}, which provide additional internal forces to 
be added in \eqref{eqn:internalf}. The strength of this approach is its 
simplicity: due to the locality of the energy, internal forces are 
straightforward to compute, and the resulting algorithm is suited for highly 
parallel implementations, including large scale GPU simulations.

However, the energy is assigned edge by edge rather than derived from the 
continuum strain-energy density of the element. Consequently every edge 
contributes only with forces directed along the edge, whereas the continuum 
element energy $U(\mathbf{F})$ couples all the edges of an element through the 
element deformation gradient, producing the full element stiffness that the 
independent springs cannot reproduce. Furthermore, from the constitutive 
standpoint, the per-edge energy is quadratic in the edge stretch, so the 
resulting stress--strain response is intrinsically linear in the small-strain 
regime and cannot reproduce the nonlinear, anisotropic stiffening exhibited by
hyperelastic soft tissues such as the myocardium
\cite{guccione1995finite,holzapfel2009constitutive}, regardless of how the 
stiffnesses $k_s$ are chosen. Extensions to nonlinear materials have been 
proposed, such as bilinear or nonlinear springs 
\cite{hammer2011mass,viola2023high}; these capture a stiffening response but, 
since the energy remains a sum of independent per-edge contributions, the coupled 
element energy is still not recovered.

\section{Consistent formulation for hyperelastic materials\label{sec:hyperel}}

The observations in section~\ref{sec:problem_springs} motivate the formulation 
introduced here. Rather than expressing the potential energy as sum of 
independent springs, we recast the continuum hyperelastic strain-energy density 
in terms of edge stretches, retaining the local, parallel structure of the 
interaction potential framework while recovering the element energy exactly. This 
change of variables has several advantages: \emph{i)} it yields interaction 
potentials energetically equivalent to continuum hyperelastic models; \emph{ii)} 
it allows nodal forces to be computed directly from edge strains; and \emph{iii)} 
it incorporates the active strain as a time evolution of the activated reference 
configuration.

\subsection{Passive mechanics}

In hyperelastic materials the internal stresses are derived from a strain-energy 
density function $\mathcal{U}$ which is usually expressed as a function of 
kinematic invariants of the right Cauchy-Green deformation tensor 
$\mathbf{C} = \mathbf{F}^T\mathbf{F}$, or the Green-Lagrange strain tensor 
$\mathbf{E} = (\mathbf{F}^T\mathbf{F} - \mathbf{I})/2$, being $\mathbf{F}$ the 
deformation gradient and $\mathbf{I}$ the identity tensor
\cite{fung2013biomechanics}. Since both $\mathbf{C}$ and $\mathbf{E}$ are 
symmetric tensors, this implies that for each control volume, only six 
independent components are required to determine the strain energy function (for 
simplicity, we use the Green-Lagrange tensor in the following). Since we adopt a 
tetrahedral discretization of the solid body, we recast the strain energy 
function as a function of the six edge strains of the tetrahedron as follows. 
For a generic edge connecting the nodes $j$ and $k$, we define the Green-Lagrange 
edge strain $\varepsilon_{jk}$ as
\begin{equation}
    \varepsilon_{jk} = \frac{1}{2}\left(\lambda^2_{jk} - 1 \right) = 
    \frac{1}{2}\left[\frac{l^2(\mathbf{x}_j, 
    \mathbf{x}_k)}{l_0^2(\mathbf{x}_j, \mathbf{x}_k)} - 1 \right]
\end{equation}
with $\lambda_{jk} = l/l_0$ the edge stretch. Given the deformation gradient 
tensor $\mathbf{F}$ of the tetrahedron, the edge in the current configuration can 
be expressed as $\mathbf{x}_{jk} = \mathbf{F}\mathbf{x}_{jk}^0$, resulting in
\begin{equation}
    \label{eqn:edgestrain}
    \varepsilon_{jk} = \frac{1}{2}\left(\frac{\mathbf{x}_{jk}^T
    \mathbf{x}_{jk}}{l_0^2} - 1 \right) = \frac{1}{2}\left(
        \frac{(\mathbf{x}_{jk}^0)^T \mathbf{F}^T \mathbf{F}
        \mathbf{x}_{jk}^0}{l_0^2} - 1 \right) =
    (\mathbf{e}_{jk}^0)^T\mathbf{E}\mathbf{e}_{jk}^0
\end{equation}
with $\mathbf{e}_{jk}^0 = (\mathbf{x}_j^0 -\mathbf{x}_k^0)/l_0$ the unit vector 
of the edge in the reference configuration. Introducing the Voigt notation 
$\mathbf{E}_v = [E_{11}, E_{22}, E_{33}, E_{12}, E_{13}, E_{23}]^T$, 
where $E_{ij}$ are the components of the Green-Lagrange tensor $\mathbf{E}$, the 
previous expression can be rewritten as 
\begin{equation}
    \varepsilon_{jk} = \mathbf{d}_{jk}^T \mathbf{E}_v
\end{equation}
where $\mathbf{d}_{jk} = [e_{jk,x}^0e_{jk,x}^0, e_{jk,y}^0e_{jk,y}^0, 
e_{jk,z}^0e_{jk,z}^0, e_{jk,x}^0e_{jk,y}^0, e_{jk,x}^0e_{jk,z}^0, 
e_{jk,y}^0e_{jk,z}^0]^T$. There are six of these equations, one for each edge, 
and the full system of equations can be written in matrix form as follows
\begin{equation}
    \label{eqn:passivecob}
    \boldsymbol{\varepsilon} = \mathbf{D} \mathbf{E}_v 
\end{equation}
where each row of the matrix $\mathbf{D}$ is given by the vector $\mathbf{d}$ 
of the corresponding edge. Provided that the tetrahedron is non-degenerate, by 
inverting this relation, we obtain a change of variables from 
$\mathbf{E}_v$ to $\boldsymbol{\varepsilon}$
\begin{equation}
    \label{eqn:cob}
    \mathbf{E}_v = \mathbf{D}^{-1} \boldsymbol{\varepsilon}
\end{equation}
which can be inserted into the strain energy density function as
\begin{equation}
    \mathcal{U} = \mathcal{U}(\mathbf{E}) = 
    \mathcal{U}(\mathbf{D}^{-1} \boldsymbol{\varepsilon}).
\end{equation}

This change of variables enables the evaluation of nodal forces directly in terms 
of edge strains
\begin{equation}
    \mathbf{f}_i = -\frac{\partial U}{\partial \mathbf{x}_i} =
        -\frac{\partial \mathcal{U}}{\partial \mathbf{x}_i}V_c^0 = 
        -\frac{\partial \mathcal{U}}{\partial \boldsymbol{\varepsilon}}
        \frac{\partial \boldsymbol{\varepsilon}}{\partial \mathbf{x}_i}V_c^0
\end{equation}
with $U = \mathcal{U} V_c^0$ the strain energy, being $V_c^0$ the reference 
volume of the tetrahedron. The first partial derivative in the last expression 
depends on the specific constitutive relation, whereas the last derivative must 
take into account only the edges incident on vertex $i$ and can be expressed as
\begin{equation}
    \frac{\partial \varepsilon}{\partial \mathbf{x}_i} = 
    \frac{\partial \varepsilon}{\partial \lambda}
    \frac{\partial \lambda}{\partial l}\frac{\partial l}{\partial \mathbf{x}_i} = 
    \frac{\lambda}{l_0} \frac{\partial l}{\partial \mathbf{x}_i}
\end{equation}
with
\begin{equation}
    \frac{\partial l}{\partial \mathbf{x}_i} = \frac{\partial |\mathbf{x}_j - 
    \mathbf{x}_k|}{\partial \mathbf{x}_i} = 
    \begin{cases}
        \frac{\mathbf{x}_j - \mathbf{x}_k}{|\mathbf{x}_j - \mathbf{x}_k|}, 
        \quad &\text{if} \quad i = j\\
        \frac{\mathbf{x}_k - \mathbf{x}_j}{|\mathbf{x}_j - \mathbf{x}_k|}, 
        \quad &\text{if} \quad i = k\\
        0, \quad &\text{else}.
    \end{cases}
\end{equation}

This formalism allows to easily modify the computation of nodal forces of 
classical interaction potential methods and introduces the non-linear effects 
which are neglected when considering independent spring elements. In the 
following sections, we first derive explicit expressions of the partial 
derivative of the strain energy function for different constitutive models and 
then extend the formalism in the case of active dynamics.

\subsubsection*{Neo-Hookean materials}

Given an isotropic neo-Hookean material, the strain energy density function for a 
cell $c$, can be expressed as (considering only the deviatoric part of the energy)
\begin{equation}
    \mathcal{U}_c = \frac{\mu}{2}\left(I_1 - 3\right)
\end{equation}
where $I_1 = \text{tr}\left(\mathbf{F}^T\mathbf{F}\right)$ is the first invariant 
and $\mu$ is a material property; since $\mathbf{F}^T\mathbf{F} = 2\mathbf{E} + 
\mathbf{I}$, the first invariant can be rewritten as 
$I_1 = 2 \text{tr}\left(\mathbf{E}\right) + 3$ leading to
\begin{equation}
    \mathcal{U}_c = \mu \text{tr}\left(\mathbf{E}_c\right) = 
    \mu\boldsymbol{\beta}_1^T\mathbf{E}_{v,c}
\end{equation}
with $\boldsymbol{\beta}_1 = [1, 1, 1, 0, 0, 0]^T$. By introducing the change of 
variables \eqref{eqn:cob}, the strain 
energy function becomes
\begin{equation}
    \label{eqn:neoHookean}
    \mathcal{U}_c = \mu \boldsymbol{\beta}_1^T \mathbf{D}_c^{-1} 
    \boldsymbol{\varepsilon}_c = \mu \mathbf{a}_c \boldsymbol{\varepsilon}_c 
\end{equation}
with $\mathbf{a}_c = \boldsymbol{\beta}_1^T \mathbf{D}_c^{-1}$; the strain energy 
partial derivative is
\begin{equation}
    \frac{\partial U_c}{\partial \boldsymbol{\varepsilon}} = \mu \mathbf{a}_c 
    V_c^0    
\end{equation}

\subsubsection*{Transversely isotropic materials: Guccione et al. (1996) 
constitutive relation}

A widely used constitutive relation for cardiac tissues is the transversely 
isotropic law \cite{guccione1995finite} that expresses the strain energy density 
function as
\begin{equation}
    \mathcal{U} = \frac{C}{2}\left(e^Q - 1\right)
\end{equation}
with
\begin{equation}
    Q = b_f E_{11}^2 + b_t\left(E_{22}^2 + E_{33}^2 + E_{23}^2 + E_{32}^2\right) 
    + b_{fs}\left(E_{12}^2 + E_{21}^2 + E_{13}^2 + E_{31}^2\right).
\end{equation}
$E_{ij}$ are the components of the Green-Lagrange strain tensor $\mathbf{E}$ in 
a local coordinate system aligned with the fiber directions (\emph{i.e.} 
$\mathbf{e}_1$ corresponds to the fiber direction $\mathbf{f}$, $\mathbf{e}_2$ to 
the cross-fiber direction $\mathbf{n}$ and $\mathbf{e}_3$ corresponds to the 
sheet direction $\mathbf{s}$); $C$, $b_f$, $b_t$ and $b_{fs}$ are material 
properties. Using the vector $\mathbf{E}_v$, the exponent $Q$ in each cell $c$ 
can be written as
\begin{equation}
    Q_c = \mathbf{E}_{v,c}^T \mathbf{B} \mathbf{E}_{v,c}
\end{equation}
with $\mathbf{B}$ a diagonal matrix of the stiffness coefficients in the 
fiber-oriented coordinate system. Introducing the change of variables 
\eqref{eqn:cob} we obtain
\begin{equation}
    Q_c = \boldsymbol{\varepsilon}_c^T \mathbf{D}_c^{-T} \mathbf{B} 
    \mathbf{D}_c^{-1} \boldsymbol{\varepsilon}_c = \boldsymbol{\varepsilon}_c^T 
    \mathbf{A}_c \boldsymbol{\varepsilon}_c
\end{equation}
where the matrix $\mathbf{A}_c$ depends only on the geometry of the tetrahedron 
and the material properties. The partial derivative of the strain energy is
\begin{equation}
    \frac{\partial U_c}{\partial \boldsymbol{\varepsilon}} = 
    \frac{\partial U_c}{\partial Q}\frac{\partial Q}{\partial 
    \boldsymbol{\varepsilon}} = Ce^{Q_c}\mathbf{A}_c\boldsymbol{\varepsilon}V_c^0
\end{equation}

\subsubsection*{Orthotropic materials: Holzapfel-Ogden (2009) constitutive 
relation}
To model orthotropic materials, we employ the strain energy density proposed in 
\cite{holzapfel2009constitutive} which reads
\begin{equation}
    \mathcal{U} = \frac{a}{2b}e^{b (I_1 - 3)} + \sum_{i=f,s}\frac{a_i}{2b_i}
    \left[e^{b_i\left(I_{4i} - 1\right)^2_+} - 1\right] +
    \frac{a_{fs}}{2b_{fs}}\left[e^{b_{fs}I_{8fs}^2} - 1\right]
\end{equation}
where the first term is the isotropic term, the second term is the transversely 
isotropic term, and the last term is the orthotropic one. The coefficients 
$a$, $a_f$, $a_s$, $a_{fs}$, $b$, $b_{f}$, $b_s$ and $b_{fs}$ are material 
properties (the first four with dimension of stress and the last four 
dimensionless), the invariants are defined as 
$I_1 = \operatorname{tr}(\mathbf{C})$, $I_{4f} = 
\mathbf{f}(\mathbf{C}\mathbf{f})$, $I_{4s} = \mathbf{s}(\mathbf{C}\mathbf{s})$, 
$I_{8fs} = \mathbf{f}(\mathbf{C}\mathbf{s})$ and $g(x)_{+} := \max{\{g(x), 0\}}$. 
To apply the change of variables, we rewrite the invariants as a function of
$\mathbf{E}$ (using the symbol $\mathcal{I}_i$ instead of $I_i$)

\begin{equation}
\begin{aligned}
I_1 &= \operatorname{tr}(\mathbf{C}) = \operatorname{tr}(2\mathbf{E} + \mathbf{I}) = 
2 \operatorname{tr}(\mathbf{E}) + 3 = 2 \mathcal{I}_1 + 3\\
I_{4f} &= \mathbf{f} (\mathbf{C} \mathbf{f}) = \mathbf{f} \cdot [(2\mathbf{E} + \mathbf{I}) \mathbf{f}] = 
2 \mathbf{f} \cdot (\mathbf{E} \mathbf{f}) + \mathbf{f} \cdot (\mathbf{I}\mathbf{f})
 = 2 \mathcal{I}_{4f} + 1\\
I_{4s} &= \mathbf{s} (\mathbf{C} \mathbf{s}) = \mathbf{s} \cdot [(2\mathbf{E} + \mathbf{I}) \mathbf{s}] = 
2 \mathbf{s} \cdot (\mathbf{E} \mathbf{s}) + \mathbf{s} \cdot (\mathbf{I}\mathbf{s})
 = 2 \mathcal{I}_{4s} + 1\\
I_{8fs} &= \mathbf{f} (\mathbf{C} \mathbf{s}) = \mathbf{f} \cdot [(2\mathbf{E} + \mathbf{I}) \mathbf{s}] = 
2 \mathbf{f} \cdot (\mathbf{E} \mathbf{s}) + \mathbf{f} \cdot (\mathbf{I}\mathbf{s})
 = 2 \mathcal{I}_{8fs}
\end{aligned}
\end{equation}
and using \eqref{eqn:cob}
\begin{equation}
    \begin{aligned}
        \mathcal{I}_1 &= \operatorname{tr}(\mathbf{E}) = \boldsymbol{\beta}_1^T\mathbf{E}_v = 
                            \boldsymbol{\beta}_1^T\mathbf{D}^{-1}\boldsymbol{\varepsilon}\\
        \mathcal{I}_{4f} &= \mathbf{f} \cdot (\mathbf{E} \mathbf{f}) = \boldsymbol{\beta}_4(f)^T \mathbf{E}_v= 
                            \boldsymbol{\beta}_4(f)^T\mathbf{D}^{-1}\boldsymbol{\varepsilon}\\
        \mathcal{I}_{4s} &= \mathbf{s} \cdot (\mathbf{E} \mathbf{s}) = \boldsymbol{\beta}_4(s)^T \mathbf{E}_v= 
                            \boldsymbol{\beta}_4(s)^T\mathbf{D}^{-1}\boldsymbol{\varepsilon}\\
        \mathcal{I}_{4fs} &= \mathbf{f} \cdot (\mathbf{E} \mathbf{s}) = \boldsymbol{\beta}_8(f,s)^T \mathbf{E}_v= 
                            \boldsymbol{\beta}_8(f,s)^T\mathbf{D}^{-1}\boldsymbol{\varepsilon}.
    \end{aligned}
\end{equation}

The strain energy density becomes
\begin{equation}
    \mathcal{U} = \frac{a}{2b}e^{2b\mathcal{I}_1 } + \sum_{i=f,s}\frac{a_i}{2b_i}
    \left[e^{4b_i (\mathcal{I}_{4i})^2_+} - 1\right]
    + \frac{a_{fs}}{2b_{fs}}\left[e^{4b_{fs}\mathcal{I}_{8fs}^2} - 1\right].
\end{equation}
and its partial derivative with respect to edge strain can be decomposed into
\begin{equation}
    \frac{\partial \mathcal{U}}{\partial \boldsymbol{\varepsilon}} = 
    \frac{\partial \mathcal{U}}{\partial \mathcal{I}_1}
    \frac{\partial \mathcal{I}_1}{\partial \boldsymbol{\varepsilon}} +
    \frac{\partial \mathcal{U}}{\partial \mathcal{I}_{4f}}
    \frac{\partial \mathcal{I}_{4f}}{\partial \boldsymbol{\varepsilon}} +
    \frac{\partial \mathcal{U}}{\partial \mathcal{I}_{4s}}
    \frac{\partial \mathcal{I}_{4s}}{\partial \boldsymbol{\varepsilon}} +
    \frac{\partial \mathcal{U}}{\partial \mathcal{I}_{8fs}}
    \frac{\partial \mathcal{I}_{8fs}}{\partial \boldsymbol{\varepsilon}}
\end{equation}
with
\begin{equation}
    \begin{aligned}
        \frac{\partial \mathcal{U}}{\partial \mathcal{I}_1} &= ae^{2b\mathcal{I}_1}, \quad 
        \frac{\partial \mathcal{U}}{\partial \mathcal{I}_{4f}} = 4a_f\mathcal{I}_{4f} e^{4b_f\mathcal{I}_{4f}^2}, \quad
        \frac{\partial \mathcal{U}}{\partial \mathcal{I}_{4s}} = 4a_s\mathcal{I}_{4s} e^{4b_s\mathcal{I}_{4s}^2}, \quad
        \frac{\partial \mathcal{U}}{\partial \mathcal{I}_{8fs}} = 4a_{fs}\mathcal{I}_{8fs} e^{4b_{fs}\mathcal{I}_{8fs}^2}\\
        \frac{\partial \mathcal{I}_1}{\partial \boldsymbol{\varepsilon}} &= \beta_1^T\mathbf{D}^{-1}, \quad
        \frac{\partial \mathcal{I}_{4f}}{\partial \boldsymbol{\varepsilon}} = \beta_4(\mathbf{f})^T\mathbf{D}^{-1}, \quad
        \frac{\partial \mathcal{I}_{4s}}{\partial  \boldsymbol{\varepsilon}} = \beta_4(\mathbf{s})^T\mathbf{D}^{-1}, \quad
        \frac{\partial \mathcal{I}_{8fs}}{\partial \boldsymbol{\varepsilon}} = \beta_8(\mathbf{f}, \mathbf{s})^T\mathbf{D}^{-1}.
    \end{aligned}
\end{equation}

\subsection{Active mechanics\label{sec:active}}
For passive dynamics, the matrix $\mathbf{D}$ can be inverted once at the 
beginning of the simulation and stored for later use. Hence, at each time step, 
one only needs to assemble the nodal forces using the appropriate constitutive 
model. The scenario changes in case of active tissues, such as the myocardium, 
since active contraction must be taken into account. We model active contraction 
using the active strain method, which employs a multiplicative decomposition of 
the deformation gradient $\mathbf{F}$ into an elastic $\mathbf{F}_e$ and an 
active contribution $\mathbf{F}_a$ \cite{ambrosi2011electromechanical}:
$\mathbf{F} = \mathbf{F}_e\mathbf{F}_a$.
The strain energy density function, then, is defined only on the basis of the 
elastic deformation gradient $\mathcal{U} = \mathcal{U}(\mathbf{F}_e^T
\mathbf{F}_e) = \mathcal{U}(\mathbf{C}_e)$, where $\mathbf{C}_e$ is the elastic 
right Cauchy-Green tensor. Here, we apply the multiplicative decomposition 
directly to the edge stretch
\begin{equation}
    \lambda = \frac{l}{l_0} = \frac{l}{l_a}\frac{l_a}{l_0} = \lambda_e \lambda_a
\end{equation}
where $l_a$ is the reference length of the edge in the active configuration, and 
the elastic Green-Lagrange edge strain $\varepsilon_e$ can be written as
\begin{equation}
    \varepsilon_e = \frac{1}{2}(\lambda_e^2 - 1) = \frac{1}{2}\left(
        \frac{(\mathbf{x}_{jk}^0)^T\mathbf{F}^T 
        \mathbf{F}\mathbf{x}_{jk}^0}{l_a^2} - 1\right) = 
    \frac{1}{2}\left(\frac{(\mathbf{x}_{jk}^0)^T\mathbf{F}_a^T\mathbf{F}_e^T
    \mathbf{F}_e \mathbf{F}_a\mathbf{x}_{jk}^0}{l_a^2} - 1\right),
\end{equation}
which is similar to that of the passive case \eqref{eqn:edgestrain}, but with the 
multiplicative decomposition applied to $\mathbf{F}$ and $\lambda$. Defining the 
elastic Green-Lagrange tensor as $\mathbf{E}_e = (\mathbf{C}_e - \mathbf{I})/2$ 
and introducing the activated reference edge as $\mathbf{x}_{jk}^a = \mathbf{F}_a 
\mathbf{x}_{jk}^0$ we can rewrite the elastic edge strain as
\begin{equation}
    \varepsilon_e = \frac{1}{2}\left(\frac{(\mathbf{x}^{jk}_a)^T\mathbf{C}_e
    \mathbf{x}^{jk}_a}{l_a^2} - 1\right) = \frac{1}{2}\left(\frac{(
        \mathbf{x}^{jk}_a)^T(2\mathbf{E}_e + \mathbf{I})\mathbf{x}^{jk}_a}{l_a^2}
         - 1\right) = (\mathbf{e}_{jk}^a)^T\mathbf{E}_e\mathbf{e}_{jk}^a
\end{equation}
with $\mathbf{e}_{jk}^a = \mathbf{x}_{jk}^a/l_a$. The change of variables now is 
performed between the elastic Green-Lagrange vector $\mathbf{E}_{e,v}$ and the 
elastic strain vector $\boldsymbol{\varepsilon}_e$
\begin{equation}
    \boldsymbol{\varepsilon}_e = \mathbf{D}_a \mathbf{E}_{e,v}
\end{equation}
which is equivalent to \eqref{eqn:passivecob} with the matrix $\mathbf{D}_a$ now 
evaluated in the active configuration. The strain energy density is evaluated as
\begin{equation}
    \mathcal{U} = \mathcal{U}(\mathbf{E}_{e,v}) = \mathcal{U}(\mathbf{D}_a^{-1} 
    \boldsymbol{\varepsilon}_e)
\end{equation}
and nodal forces are given as
\begin{equation}
    \mathbf{f}_i = -\frac{\partial U}{\partial \mathbf{x}_i} = 
    -\sum_{c \in \mathcal{E}_c^i} \frac{\partial \mathcal{U}_c}
    {\partial \mathbf{x}_i}V_{c}^0 = -\sum_{c \in \mathcal{E}_c^i}\sum_{e \in 
    \mathcal{E}_{c,\ell}^i} \frac{\partial \mathcal{U}_c}{\partial 
    \varepsilon_{e,\ell}}\frac{\partial \varepsilon_{e,\ell}}{\partial 
    \mathbf{x}_i}V_{c}^0
\end{equation}
with
\begin{equation}
\frac{\partial \varepsilon_e}{\partial \mathbf{x}_i} =
\frac{\partial \varepsilon_e}{\partial \lambda_e}\frac{\partial \lambda_e}
{\partial \mathbf{x}_i} = \frac{\partial \varepsilon}{\partial \lambda_e}
\frac{\partial \lambda_e}{\partial l}\frac{\partial l}{\partial \mathbf{x}_i} = 
\frac{\lambda_e}{l_a}\frac{\partial l}{\partial \mathbf{x}_i}.
\end{equation}

It is worth noticing that, the reference volume $V_c^0$ in the active dynamic is 
equal to the reference volume of the passive configuration only if the active 
deformation is incompressible, \emph{i.e.} $\operatorname{det}(\mathbf{F}_a) = 1$. 
Otherwise, the reference volume should be replaced with the corresponding value
of the active reference configuration. In active dynamics, the $6 \times 6$ 
matrix inversion $\mathbf{D}^{-1}$ must be performed each timestep, since the 
reference configuration changes with the activation of the tissue. This step can
be performed efficiently in parallel, as explained in section \ref{sec:numerics}.

\subsection{Mechanical energy balance}
To verify that the present formulation preserves the mechanical energy balance, 
we derive the corresponding conservation equation by multiplying the dynamical 
equation \eqref{eqn:Newton} for each node by its velocity and summing over all 
nodes:

\begin{equation}
    \label{eqn:kin1}
    \sum_{i = 1}^{N_v} m_i \mathbf{v}_i \cdot \frac{\mathbf{dv}_i}{dt} = 
    \sum_{i = 1}^{N_v} \mathbf{f}_i \cdot \mathbf{v}_i = -\sum_{i = 1}^{N_v} 
    \left(\sum_{c \in \mathcal{E}_c^i}
    \frac{\partial U_c}{\partial \mathbf{x}_i}\right) \cdot \mathbf{v}_i
    = -\sum_{i = 1}^{N_v} 
    \left(\sum_{c \in \mathcal{E}_c^i}
    \frac{\partial U_c}{\partial \boldsymbol{\varepsilon}_c}
    \frac{\partial \boldsymbol{\varepsilon}_c}{\partial \mathbf{x}_i}\right) 
    \cdot \mathbf{v}_i
\end{equation}
where $\mathbf{v}_i = d\mathbf{x}_i / dt$, and we have assumed that 
there is no viscous dumping ($\xi = 0$) and the absence of external forces 
($\mathbf{f}_i^{ext} = 0$). The left-hand side is the time derivative 
of the kinetic energy
\begin{equation}
    \sum_{i = 1}^{N_v} m_i \mathbf{v}_i \cdot \frac{\mathbf{dv}_i}{dt} =
    \frac{d}{dt}\left(\sum_{i = 1}^{N_v} \frac{1}{2}m_i |\mathbf{v}_i|^2 \right) 
    = \frac{dK}{dt}.
\end{equation}
The right-hand side can be rewritten by dividing the total differential of 
the strain energy by $dt$
\begin{equation}
    \frac{dU}{dt} = \sum_{c = 1}^{N_c} \frac{dU_c}{dt} = \sum_{c = 1}^{N_c}
    \left(\frac{\partial U_c}{\partial \mathbf{D}^{-1}_c}: \frac{d
    \mathbf{D}_c^{-1}}{dt} +
    \frac{\partial U_c}{\partial \boldsymbol{\varepsilon}_c}\cdot 
    \frac{d\boldsymbol{\varepsilon}_c}{dt}\right)
\end{equation}
and, noticing that $\boldsymbol{\varepsilon}(\mathbf{x},\mathbf{x}^0)$,
the last term can be expanded as
\begin{equation}
    \frac{d\boldsymbol{\varepsilon}_c}{dt} = \sum_{i \in c}
    \frac{\partial\boldsymbol{\varepsilon}_c}{\partial \mathbf{x}_i}
    \frac{d \mathbf{x}_i}{dt} + \sum_{i \in c}
    \frac{\partial\boldsymbol{\varepsilon}_c}{\partial \mathbf{x}_i^0}
    \frac{d\mathbf{x}_i^0}{dt}.
\end{equation}
Therefore, we can write
\begin{equation}
    \sum_{c = 1}^{N_c}\frac{\partial U_c}{\partial \boldsymbol{\varepsilon}_c}
    \cdot 
    \left(\sum_{i \in c}
    \frac{\partial \boldsymbol{\varepsilon}_c}{\partial \mathbf{x}_i}
    \frac{d\mathbf{x}_i}{dt}\right) = 
    \frac{dU}{dt} - \sum_{c = 1}^{N_c}
    \left[ \frac{\partial U_c}{\partial \mathbf{D}^{-1}_c}: \frac{d
    \mathbf{D}_c^{-1}}{dt} + \frac{\partial U_c}{\partial 
    \boldsymbol{\varepsilon}_c}\cdot
    \left(\sum_{i \in c}\frac{\partial\boldsymbol{\varepsilon}_c}{\partial \mathbf{x}_i^0}
    \frac{d \mathbf{x}_i^0}{dt}\right)\right] = 
    \frac{dU}{dt} - \mathcal{P}_a
\end{equation}
where the last term represents the power associated with the active 
deformation. By substituting into equation \eqref{eqn:kin1} we find the 
balance equation for the mechanical energy $E$
\begin{equation}
    \label{eqn:energy}
    \frac{d(K+U)}{dt} = \frac{dE}{dt} = \mathcal{P}_a
\end{equation}

The numerical solver must ensure that in passive dynamics the mechanical energy 
is conserved (within the numerical accuracy of the time discretization), and in 
active dynamics the change in mechanical energy must balance the active work. 
This is verified numerically in section \ref{sec:validation}.

\subsection{Incompressibility constraint\label{sec:incomp}}

The simulation of incompressible deformable solids is a challenging numerical 
problem. In FEM, Lagrangian multipliers or penalty methods are employed 
\cite{bonet1997nonlinear}. These approaches introduce an additional variable, 
which can be regarded as a pressure field, subject to the 
Ladyzhenskaya–Babuska–Brezzi (LBB) condition. Violation of this condition results 
in spurious pressure modes and volumetric locking, and it is often avoided 
adopting an independent discretization for the pressure field. 

In the context of interaction potential methods, a natural strategy to enforce 
material incompressibility is the inclusion of an energy potential function on 
the cell volume:
\begin{equation}\label{eq:volcell}
    U_c = \frac{1}{2}k_v\left(\frac{V_c - V_c^0}{V_c^0} \right)^2 V_c^0 = 
    \frac{1}{2}k_v \left(J_c - 1\right)^2 V_c^0,
\end{equation}
where $J_c = \text{det}(\mathbf{F}_c)$ is the determinant of the deformation 
gradient of cell $c$, and $k_v$ is an elastic stiffness, which can be regarded as 
the bulk modulus of the material. However, since the number of cells in a 
tetrahedral mesh can be order $4 N_v$, whereas the degrees of freedom are 
$3 N_v$, adding the volumetric constraint~\eqref{eq:volcell} for each cell makes 
the system highly overconstrained, thus resulting in volumetric locking 
\cite{irving2007volume,francu2021locking}. To overcome this issues (and without 
the introduction of an additional pressure field) we propose to rewrite the 
volume potential in terms of nodal volume instead, $\emph{i.e.}$ for each node 
we define a control volume
\begin{equation}
    V_i = \sum_{c \in \mathcal{E}_c^i} \frac{V_c}{4}
\end{equation}
and express the volume energy as
\begin{equation}
    U_i = \frac{1}{2}k_v \left(\frac{V_i - V_i^0}{V_i^0} \right)^2 V_i^0
\end{equation}
which results in only $N_v$ constraint, thus avoiding volumetric locking. In 
section \ref{sec:validation} we demonstrate that this formulation converges to 
the incompressible solution increasing the volumetric stiffness $k_v$ and that 
the error on the incompressibility constraint decays first order with $k_v$.

\section{Numerical implementation\label{sec:numerics}}

The proposed method is implemented in an in-house code for Fluid-Structure 
Interaction simulations which has already been used and validated for 
biomechanical problems \cite{viola2023high}. The software is a multi-architecture 
code for CPUs and GPUs written in Fortran and CUDA. Equation \eqref{eqn:Newton} 
is solved with a semi-implicit Euler method, which is an explicit first order in 
time and symplectic method. Matrix inversions for $\mathbf{D}$ \eqref{eqn:cob} 
are performed using the utility \emph{cublasDmatinvBatched} from the 
\textbf{cublas} library (https://docs.nvidia.com/cuda/cublas/index.html). 
Differently than the common approach used in FEM solvers for active dynamics 
simulations \cite{rossi2014thermodynamically,goktepe2014generalized,
garcia2019new,propp2019orthotropic}, in which the numerical solution requires to 
build a global stiffness matrix and to iteratively solve for the quasi-static or 
dynamic problem, this framework provides a fast alternative, being highly 
parallelized, and suitable for large-scale GPU simulations. The structure of the 
timestep is the following:
\begin{enumerate}
    \item update the reference configuration due to the active deformation 
    gradient $\mathbf{F}_a$ and perform matrix inversion $\mathbf{D}^{-1}$ 
    in each cell;
    \item compute internal forces on each node due to each interaction potential 
    using \ref{eqn:internalf};
    \item update the position vector of each node using \eqref{eqn:Newton};
    \item enforce boundary conditions.    
\end{enumerate}
Step 1 is performed only in active deformation. Step 1 and step 2 scale with the 
number of cells, hence their computational cost is $\mathcal{O}(N_c)$; step 3 
scales with the number of nodes and its computational cost is $\mathcal{O}(N_v)$. 
Since all computations are local, they can be easily made parallel, although
concurrent updates of shared nodes requires appropriate accumulations strategies 
(as atomic adds on GPU). For the simulation performed in section 
\ref{sec:ventricle}, which employs a computational mesh having 1885262 cells, the
average time per timestep is $21.2$ ms on a NVIDIA A100 80 GB PCIe GPU, with 
a total computation time of 30 minutes. 

The implementation of the proposed method, including the structural GPU solver, 
is openly available at \url{https://gitlab.com/gssi-fluids/iphem}.

\section{Validation and benchmark problems\label{sec:validation}}

In this section, we present several test cases to validate both passive and 
active dynamics using different constitutive relations. We start by validating 
the passive dynamics, solving for the deformation of a beam and the inflation of 
an idealized ventricle using the transversely isotropic model; then we validate 
the orthotropic model in both passive and active configurations, solving for the 
passive compression of a cube and the active deformation of a cube with a 
prescribed fiber shortening time history. Finally, we apply the proposed method 
to a human ventricle prescribing the activation time history. Since we aim to 
present a robust validation of the mechanical solver, we have decoupled the 
mechanical deformation from the electrical problem, making the results 
independent from the underlying electrophysiology solver and thus more 
reproducible.

\subsection{Deformation of a beam\label{sec:beam}}

\begin{figure}[!h]
	\centering
	\includegraphics[width=\textwidth]{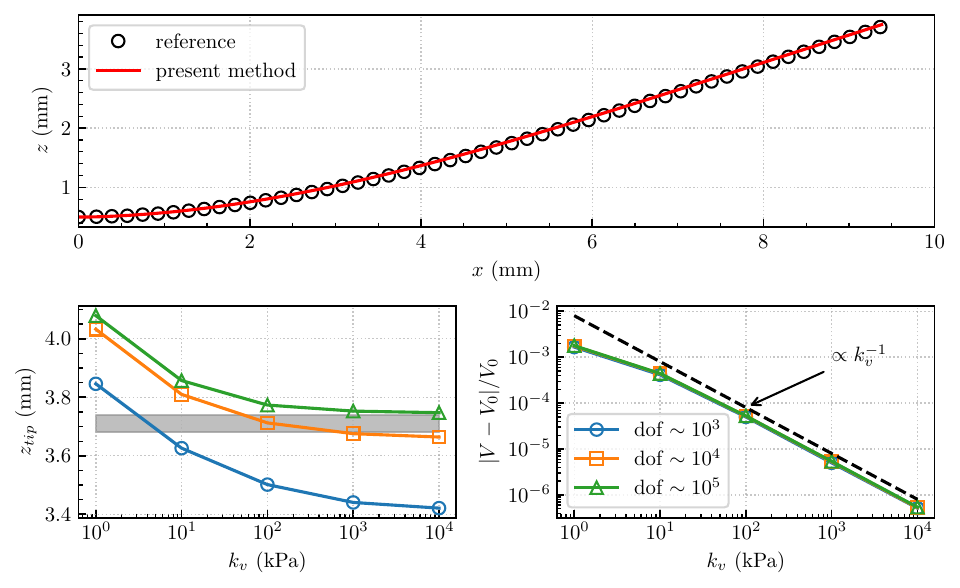}
	\caption{Results for the beam deflection test case: (top panel) shape of the 
    deformed fiber initially at $(x, 0.5, 0.5)$; (bottom left) vertical position 
    of the beam tip $z_{tip}$ as function of bulk modulus $k_v$ and number of 
    degree of freedoms (the shaded gray area represent the range of literature 
    data); (lower right) convergence rate of the incompressibility constraint 
    with the volumetric stiffness $k_v$.}
	\label{fig:fig1}
\end{figure}

Following \cite{land2015verification}, we solve for the deformation of a 
rectangular beam under a uniform pressure load. The undeformed geometry is the 
region $x \in [0, 10], y \in [0,1], z \in [0,1]$ mm; the solid is modeled as 
transversely isotropic with $C = 2$ kPa, $b_f = 8$, $b_t = 2$ and $b_{fs} = 2$ 
and the fibers are directed along the main beam axis, \emph{i.e.} $\mathbf{f} = 
(1,0,0)$. Dirichlet boundary conditions are applied at $x = 0$, 
hence for all nodes belonging to the left face of the beam, we enforce
\begin{equation}
    \mathbf{x}_i(t) = \mathbf{x}_i(t=0), \quad \mathbf{v}_i(t) = 0, \quad t > 0.
\end{equation}
At $z=0$, a uniform pressure equal to $P = 0.004$ kPa is applied; the 
corresponding nodal forces are obtained evaluating for each triangular face $t$ 
of the beam surface, the pressure force $\mathbf{f}_t^{ext} = -P S 
\widehat{\mathbf{n}}$, where $S$ is the surface of the triangle, and 
$\widehat{\mathbf{n}}$ the unit normal vector pointing away from the surface. 
This force is then equally distributed to the nodes belonging to triangle $t$ and 
the resulting external force acting on node $i$ is 
\begin{equation}
    \mathbf{f}_i^{ext} = \sum_{t \in \mathcal{E}_t^i} \frac{\mathbf{f}_t^{ext}}{3}
\end{equation}
with $\mathcal{E}_t^i$ the set of all faces incident on vertex $i$. The results 
for the test case are reported in Figure \ref{fig:fig1}: the top panel displays 
the deformed location of the line $(x, 0.5, 0.5)$, compared with results from 
\cite{land2015verification}; bottom left panel displays the vertical position of 
the tip of the previously defined line, for different values of volumetric 
stiffness $k_v$ and number of degrees of freedom, while, in the bottom right 
panels the error of the incompressibility constraint is reported for the same 
parametric space. The proposed method provides a deformation profile in good 
agreement with the literature results, and it converges to the incompressible 
solution for the volumetric stiffness going to infinity without volumetric 
locking, with the global error on the incompressibility constraint decaying first 
order with the bulk modulus.

\subsection{Inflation of a ventricle}

\begin{figure}[!h]
	\centering
	\includegraphics[width=0.8\textwidth]{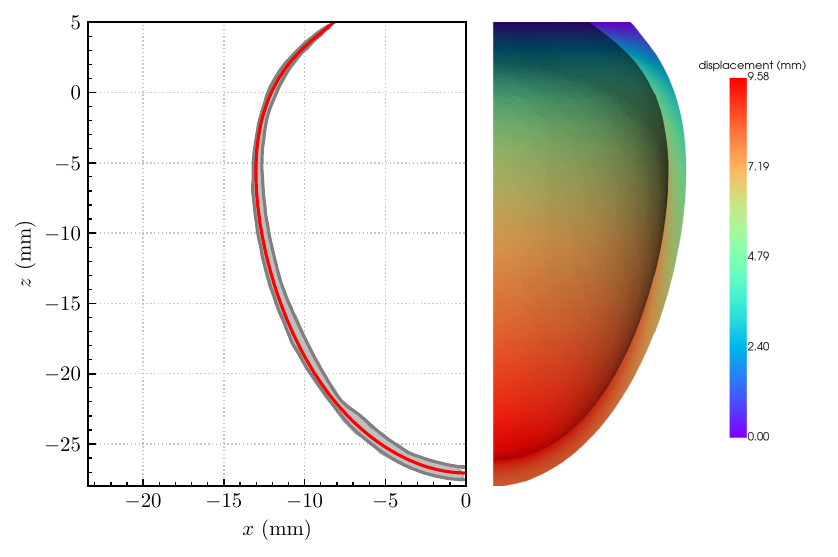}
	\caption{Results for the ventricle inflation test case: the left half of the 
        plot report the deformation of the sampled fiber (solid red line 
        \textcolor{red}{\rule[0.5ex]{1em}{1pt}}) with the range of literature 
        results highlighted in gray \cite{land2015verification}; the right half 
        of the plot display the deformed ventricle with colors representing the 
        displacement field.}
	\label{fig:fig2}
\end{figure}

Next, we solve for the inflation of a ventricle as proposed in 
\cite{land2015verification}. The undeformed geometry is a truncated ellipsoid 
given by the following parametrization:
\begin{equation}
    \mathbf{x} = 
    \begin{pmatrix} x \\ y \\ z \\ \end{pmatrix} = 
    \begin{pmatrix} r_s \sin{u} \cos{v} \\ r_s \sin{u} \sin{v} \\ r_l \cos{u} 
    \end{pmatrix}
\end{equation}
with the solid enclosed by three surfaces:
\begin{itemize}
\item[-] the \emph{endocardial surface}: $r_s = 7$ mm, $r_l = 17$ mm, $u 
\in [-\pi, -\arccos \frac{5}{17}], v \in [-\pi, \pi]$, 
\item[-] the \emph{epicardial surface}: $r_s = 10$ mm, $r_l = 20$ mm, $u 
\in [-\pi, -\arccos \frac{5}{20}], v \in [-\pi, \pi]$,
\item[-] the \emph{base plane}: $z = 5$ mm.
\end{itemize}
The material is assumed to be isotropic, with material parameters $C = 10$ kPa, 
$b_f = b_t = b_{fs} = 1$. Dirichlet boundary conditions on position and velocity 
are applied at the base plane and a uniform pressure load with $P = 10$ kPa is 
applied on the endocardial surface. The test has been carried out with a mesh 
that has 615343 tetrahedra and 116371 nodes; the incompressibility constraint is 
imposed using a stiffness parameter $k_v = 10$ MPa, resulting in a maximum 
relative variation of the ventricle volume around $10^{-4}$. Figure 
\ref{fig:fig2} shows the deformed ventricle (right half) with the deformed shape 
of the fiber located in the middle of the ventricular wall ($8.5 \sin{u}, 0, 
18.5 \cos{v}$) highlighted in red (left half). The results fall well within the
range of data from the literature \cite{land2015verification} .

\subsection{Compression of a cube}

Following \cite{pezzuto2014orthotropic}, we solve for the compression of a unit 
cube modeled as an orthotropic material. The solid is defined in the domain 
$\Omega_0 = [0,1] \times [0,1] \times [0,1]$ mm, with fibers oriented as 
$\mathbf{f} = \cos{\alpha} \mathbf{e}_1 + \sin{\alpha} \mathbf{e}_2$, and 
uniformly distributed in direction $\mathbf{e}_3$. Uniform external pressure is 
applied at the boundaries $z = 0$ mm and $z = 1$ mm, while at $y = 0$ mm we 
enforce zero displacement, velocity, and acceleration in direction $\mathbf{e}_2$. 
For this particular setup, the deformation is a combination of axial strain and 
simple shear in the form of
\begin{equation}
    \mathbf{F} = 
        \begin{bmatrix}
            \lambda_1 & \kappa \lambda_2 & 0\\
            0 & \lambda_2 & 0\\
            0 & 0 & \lambda_3
        \end{bmatrix}
\end{equation}
and an analytical expression for the invariants can be derived
\begin{equation}
    \label{eqn:Invariants}
    \begin{aligned}
        I_1 &= \lambda_1^2 + (\kappa^2 + 1)\lambda_2^2 + (\lambda_1 \lambda_2)^{-2}\\
        I_{4f} &= (\lambda_1 \cos{\alpha} + \kappa \lambda_2 \sin{\alpha})^2 + 
        \lambda_2^2 \sin^2{\alpha}.
    \end{aligned}
\end{equation}
Note that we report only two invariants since, for this test case, 
$I_{4s} = I_{8fs} \equiv 0$.
We perform a numerical simulation up to steady state, defined as the time step 
after which relative changes in potential energy and the absolute value of the 
kinetic energy are smaller than a numerical tolerance $\epsilon = 10^{-12}$.
All material properties and numerical parameters are listed in table 
\ref{tab:tab1}.
\begin{table}
	\caption{Material properties and numerical setup for the cube compression 
    test case.}
	\centering
	\begin{tabular}[c]{|c|c|c|c|}
        \hline
        $a$ & $a_f$ & $a_s$ & $a_{fs}$ \\
        \hline
        0.333 kPa & 18.535 kPa & 2.564 kPa & 0.417 kPa \\
        \hline
        $b$ & $b_f$ & $b_s$ & $b_{fs}$ \\
        \hline
        9.242 & 15.972 & 10.446 & 11.602 \\
        \hline        
        $P$ & $k_v$ & $\alpha$ & $\xi / \rho$ \\
        \hline
        $1$ kPa & $10^5$ kPa & $45^{\circ}$ & $10^{-4}$ m\textsuperscript{3}/s \\
        \hline
    \end{tabular}
	\label{tab:tab1}
\end{table}

From the steady state solution and the initial configuration, we solve for the 
deformation gradient $\mathbf{F}$ using a least-square method. Then we evaluate 
the error between the numerical invariants and the analytical values given in 
equations \eqref{eqn:Invariants} and report the $L_{\infty}$ and the $L_2$ norms 
of the error vectors in table \ref{tab:tab2}, alongside the non-null values of 
the deformation gradient. The proposed numerical solver accurately evaluates the 
invariants; it is worth noticing that the determinant of $\mathbf{F}$ is 
$\operatorname{det}(\mathbf{F}) = 0.99998115$ which is consistent with the change 
in the total volume computed from the simulation, which is order $10^{-5}$.

\begin{table}
	\caption{Results for the cube compression test case.}
	\centering
	\begin{tabular}[c]{|c|c|c|c|}
        \hline
        $\lambda_1$ & $\lambda_2$ & $\lambda_3$ & $k$ \\
        \hline
        1.111014 & 1.0933259 & 0.82323281 & -0.18060284 \\
        \hline
        $L_{\infty}(I_1)$ & $L_{2}(I_1)$ & $L_{\infty}(I_{4f})$ & $L_{2}(I_{4f})$ \\
        \hline
        $8.1767811e-06$ & $8.0326326e-06$ & $2.8848066e-07$ & $9.5521618e-08$ \\
        \hline        
    \end{tabular}
	\label{tab:tab2}
\end{table}

\subsection{Active deformation of a cube}

\begin{figure}
    \centering
    \includegraphics[width=\textwidth]{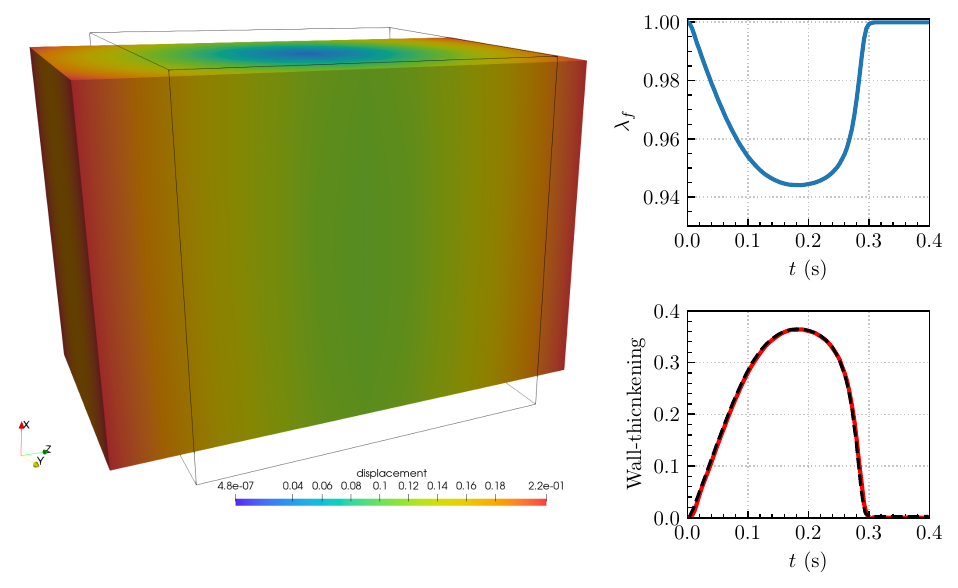}
    \caption{(left column) Cube configuration at maximum contraction with 
    colorbar showing the displacement field in each tetrahedron. (right column) 
    Imposed fiber stretching time history (top panel); time history of the wall 
    thickening (bottom panel): (\textcolor{red}{--}) present results, (- -) 
    reference results from \cite{rossi2014thermodynamically}.
    \label{fig:fig3}}
\end{figure}

In this test case, we solve for the active deformation of the same cube of the 
previous test case, as in \cite{rossi2014thermodynamically}. In the original test 
case, the mechanical deformation is coupled with the diffusion of the action 
potential; however, since the propagation of the depolarization front is much 
faster than the mechanical response, it can be assumed that the deformation field 
$\lambda_f$, along the fiber direction, is uniform inside the solid. Hence, for 
the time history of $\lambda_f$, we prescribe the average value reported in 
Figure 5 of \cite{rossi2014thermodynamically}. The active stretch in the other
two direction are given by
$$
\lambda_n = k_0 \lambda_f, \quad \lambda_s = \frac{1}{\lambda_f \lambda_n}
$$
which completely determine the active deformation gradient $\mathbf{F}_a$
and enforce the inompressibility constraint $\operatorname{det}(\mathbf{F}_a) = 1$. 
In this test case, we used $k_0 = 4$. In Figure \ref{fig:fig3} (left column) we 
report the geometry of the cube at 
maximum contraction, alongside the bounding box of the initial configuration; 
we also display (right column) the 
prescribed time evolution of the fiber stretching (top panel) and a comparison of 
the wall thickening between the present solver and the results of 
\cite{rossi2014thermodynamically}, which are in excellent agreement.

\subsection{Energy balance}

\begin{figure}
    \centering
    \includegraphics[width=\textwidth]{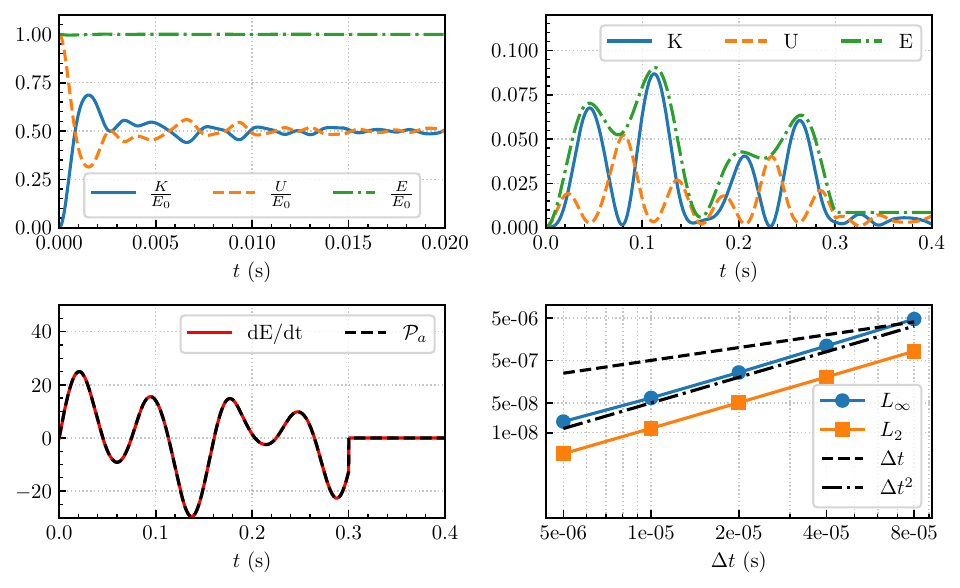}
    \caption{(upper left) Time history of kinetic ($K/E_0$), potential ($U/E_0$) 
    and mechanical ($E/E_0$) energy, normalized with the initial mechanical 
    energy ($E_0 = E(t=0)$, with $E = K + U$) for the passive dynamics test case. 
    (upper right) Time history of kinetic ($K$), potential ($U$) and mechanical 
    ($E$) energy for the active dynamics test case. (lower left) Comparison 
    between left-hand side ($dE/dt$) and right-hand side ($\mathcal{P}_a$) of 
    equation \eqref{eqn:energy}. (lower right) Time convergence rate of the 
    $L_{\infty}$ and $L_2$ error norms of equation \eqref{eqn:res}.
    \label{fig:fig4}}
\end{figure}

To verify that the proposed method satisfies the mechanical energy balance, 
we perform two simulations, one for passive dynamics and one for
active dynamics. In both simulations, the solid domain is a unit cube 
with edge length $1$ mm, modeled with the transversely isotropic constitutive
law and material parameters as in the test case \ref{sec:beam}. 
In the passive test case, we apply a random displacement to the
initial node location with amplitude equal to 1\% of the mean edge length of the 
mesh, applied independently for each component of the position vector. We track 
in time the kinetic energy and the potential energy, which are reported in figure 
\ref{fig:fig4} (upper left panel), together  with the 
mechanical energy. All quantities are normalized by the initial mechanical 
energy value $E(t=0)$. The non-dimensional mechanical energy is equal to 1 and, 
after an initial transient, energy equipartition is reached, with constant 
exchange between kinetic and potential energy, which oscillate around 0.5. In the 
active case, there is no mechanical energy conservation, and the time variation 
of the energy must be balanced by the active power injected or extracted by the 
fibers $\mathcal{P}_a$. For the same unit cube, we impose the time history of the 
fiber contraction as 
\begin{equation}
    \lambda_f = \begin{cases}
        1 - a\sin\left(\frac{2 \pi t}{T}\right), \quad &\text{if} \quad t \le T \\
        1 &\text{else}
    \end{cases}
\end{equation}
with $a = 0.05$ and $T = 0.3$. The time history of the kinetic, potential, and 
mechanical energy is reported in figure \ref{fig:fig4} (upper right panel), 
highlighting that the mechanical energy varies in time during the active phase,
whereas it is conserved for $t > T$. Since the active power represents a change 
of the strain energy due to a change of the reference configuration, it 
corresponds to the time derivative of the total strain energy at fixed position; 
therefore, we numerically approximate the active power as
\begin{equation}
    \mathcal{P}_a^n = \frac{dU^n}{dt}\Big|_{\mathbf{x}} = \frac{U(\mathbf{x}^n,
    (\mathbf{x}^0)^{n+1}) - U(\mathbf{x}^n,(\mathbf{x}^0)^{n})}{dt} + 
    \mathcal{O}(\Delta t)
\end{equation}
Figure \ref{fig:fig4} (lower left panel) displays a comparison between the 
left-hand side $dE/dt$ and the right-hand side $\mathcal{P}_a$ of equation 
\eqref{eqn:energy}. To evaluate the accuracy on energy conservation, we compute 
the error at each timestep as
\begin{equation}
    \label{eqn:res}
    R^n = |\Delta E^n - W_a^n|
\end{equation}
with $\Delta E^n = E^{n+1} - E^n$ and $W_a^n = U(\mathbf{x}^n,(
\mathbf{x}_0)^{n+1}) - U(\mathbf{x}^n,(\mathbf{x}_0)^{n}$ and report the 
$L_{\infty}$ and $L_2$ norms of the error in figure \ref{fig:fig4} (lower right 
panel), which exhibit second order convergence rate with the 
timestep.

\section{Human ventricle active dynamics\label{sec:ventricle}}

\begin{figure}
\centering
    \includegraphics[width=\textwidth]{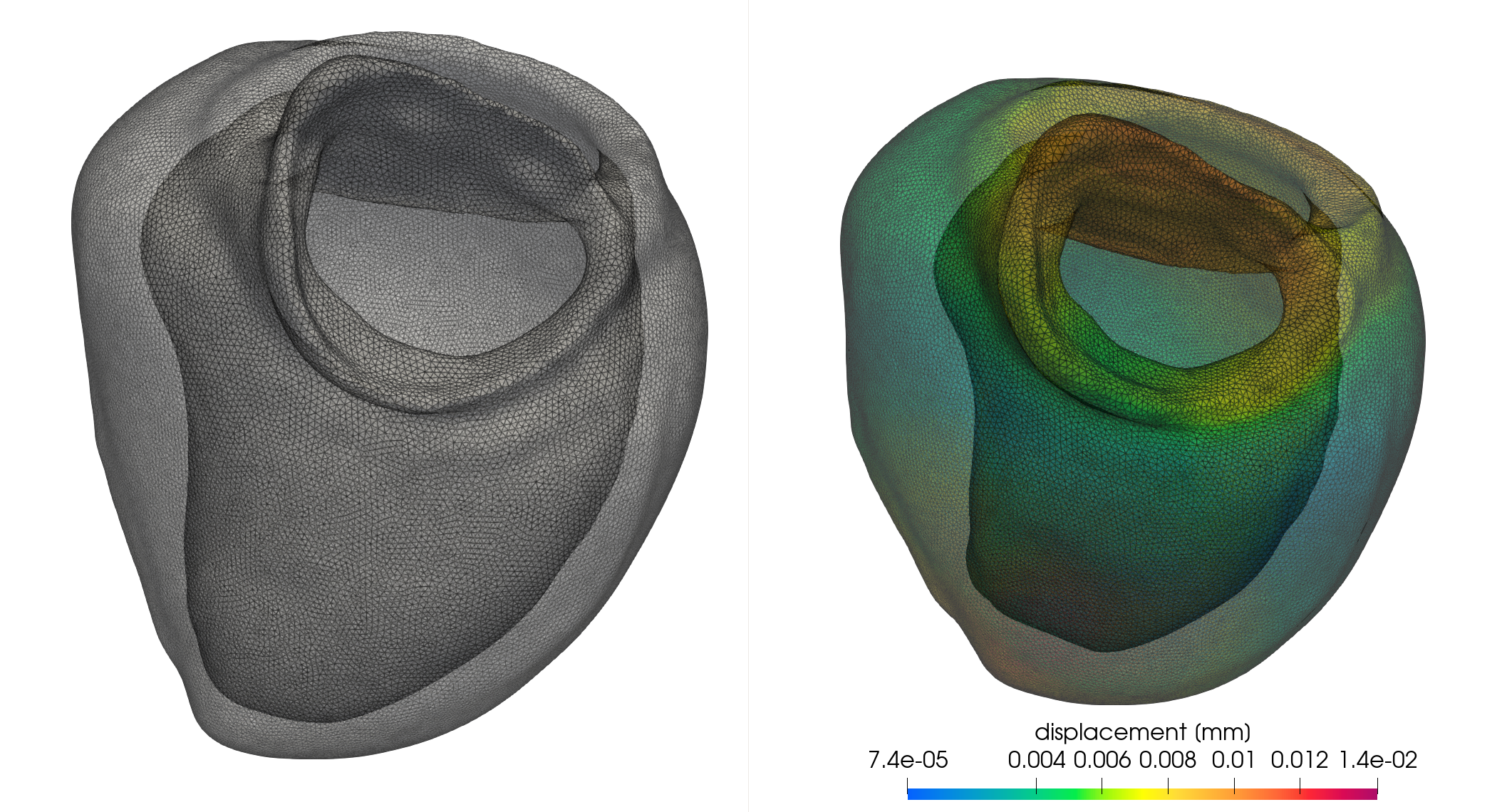}
    \caption{Realistic ventricle test case: (left panel) initial stress-free 
    configuration; (right panel) maximum contraction configuration, with colors 
    representing the displacement field.\label{fig:fig5}}
\end{figure}

\begin{figure}
\centering
    \includegraphics[width=\textwidth]{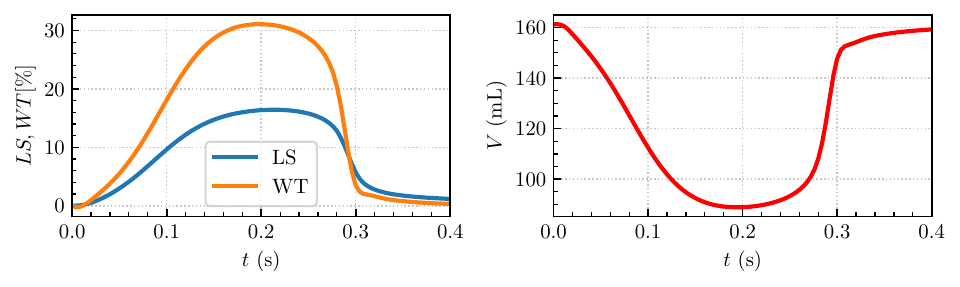}
    \caption{(left panel) Time history of the longitudinal shortening (LS) and 
    the wall thickening (WT); (right panel) time history of the endocardial 
    volume (V).\label{fig:fig6}}
\end{figure}

Finally, to demonstrate the applicability of the proposed method to more 
realistic cases, we simulate the active contraction of a realistic ventricle, 
taken from the public repository https://lifex.gitlab.io/fibers.html. The mesh 
consists of 330721 nodes and 1885262 cells; the repository also provides the 
fiber field, obtained by solving a Laplace equation (see \cite{africa2023lifex} 
for more details). The ventricle is modeled with the orthotropic constitutive law 
with material properties as in table \ref{tab:tab1} and the activation is imposed 
as in figure \ref{fig:fig3} (upper right panel), uniformly distributed in the 
solid with $k_0 = 5$. Although this will lead to a non realistic deformation of 
the ventricle, the goal is to provide an example of mechanical deformation 
independent of the cellular models used to propagate the action potential. The 
left panel of figure \ref{fig:fig5} displays the stress-free configuration of 
the ventricle, while the right panel displays the configuration at maximum 
contraction colored with the displacement field. For this simulation, the 
proposed method predicts a maximum longitudinal shortening of approximately 20\%, 
a maximum value of the average wall thickening of approximately 32\%, and an 
ejection fraction of about 45\%, which are in line with physiological 
values. The longitudinal shortening, mean wall thickening, and endocardial volume 
time histories are reported in figure \ref{fig:fig6}. 
An animation of the simulation is available online as supplementary material.

\section{Conclusion}

In this work, we have presented an edge-based interaction potential method for 
the simulation of active biological tissues. The method relies on expressing the 
strain energy density function for hyperelastic materials directly in terms of 
edge strain of a tetrahedral mesh. Since the deformation gradient of a linear
tetrahedron is recovered exactly from its edge vectors, the discrete energy is 
the continuum strain-energy density itself rather than a spring-network 
approximation. In this way, the formulation preserves the energetic structure of 
the continuum model, while expressing the internal forces as local, edge-wise 
contributions in the form natural to interaction-potential solvers. We have 
further proposed a strategy for representing nearly incompressible materials 
without volumetric locking. The proposed approach has several advantages: 
\emph{i)} it ensures energetic consistency with continuum hyperelastic model; 
\emph{ii)} it enables the natural incorporation of the active-strain 
multiplicative decomposition through a time-dependent activation of the reference 
configuration; \emph{iii)} the diagonal mass matrix and purely local force 
evaluation make it well suited to explicit, highly parallel implementations, and 
we provide a parallel GPU implementation, released as open-source 
software\footnote{\url{https://gitlab.com/gssi-fluids/iphem}}.

The solver was validated on a set of benchmark problems spanning different 
hyperelastic constitutive laws, against both reference data from the literature 
and finite element solutions, and was shown to reproduce the expected mechanical 
behavior. The analysis of the energy balance confirms that the scheme preserves 
energy in both the passive and active regimes, with a residual that decays to 
second order in time. Finally, the method was applied to the active deformation 
of a realistic ventricle, predicting longitudinal shortening and wall-thickening 
values consistent with the literature.

The present formulation is deliberately scoped to the structural problem and has 
two attendant limitations. First, it is built on linear (P1) tetrahedra: the 
exact recovery of the deformation gradient, and hence the energetic equivalence 
with the continuum, relies on the constant-strain property of the linear element 
and does not carry over unchanged to higher-order interpolations, whose extension 
we leave to future work. Second, the diagonal mass matrix and local forces are 
intended for explicit time integration, which couples the method to the usual 
stability constraint on the time step; the gain in per-step cost and parallel 
efficiency is what offsets this restriction. 

We emphasize that the absence of an electrophysiology model is not a limitation 
of the method but a matter of scope: the active strain enters the formulation as 
a prescribed activation, and coupling it to an excitation--contraction model is 
a modeling choice external to the structural solver. This last point also defines 
the natural path forward. The locality and parallel efficiency that motivate the 
method make it a candidate component for whole-heart digital twins, in which
the structural model is coupled to electrophysiology and to the surrounding blood 
flow. Our group has developed the constituent solvers for such a framework---a 
fast electrophysiology model of the whole heart \cite{delcorso2022fast}, a 
GPU-accelerated cardiac digital twin \cite{viola2023gpu}, and an 
immersed-boundary fluid--structure model of the heart \cite{viola2023high}---and 
the integration of the present formulation into this 
electrophysiology--structure--fluid pipeline, with application to 
patient-specific geometries and cardiac disease, is the principal direction of 
future work.

\section*{Code availability}
The implementation of the proposed method, including the GPU solver, is openly available at \url{https://gitlab.com/gssi-fluids/iphem}.

\section*{Acknowledgments}
This work was supported by the Italian Ministry of University and Research (MUR) 
under the PRIN 2022 project, grant no. 2022J5ZNHS\_001 (CUP: D53C24004090001).
F.V. acknowledges financial support by the European Research Council (ERC) under 
the European Union’s Horizon Europe research and innovation program, Project 
CARDIOTRIALS (Grant Agreement No. 101039657).

\bibliographystyle{plain}
\bibliography{biblio}

\end{document}